\documentclass[aps,prb,reprint]{revtex4-1}
\pdfoutput=1

\usepackage{graphpap}
\usepackage[dvips]{graphicx}
\usepackage[dvips]{graphics}
\usepackage{color}
\usepackage[normalem]{ulem}
\usepackage {ulem}
\usepackage{amsmath}
\usepackage[T1]{fontenc}
\usepackage[latin9]{inputenc}
\usepackage{array}
\usepackage{multirow}
\usepackage{amssymb}
\usepackage{textcomp}
\usepackage{soul}

\begin{document}

\title{Low $B$~field magneto-phonon resonances in single-layer and bilayer graphene}

 \author{C. Neumann$^{1,2}$, S. Reichardt$^1$, M. Dr\"ogeler$^1$, B. Terr\'es$^{1,2}$,  K.~Watanabe$^3$, T.~Taniguchi$^3$, B. Beschoten$^1$, S. V. Rotkin$^{1,4}$ and C. Stampfer$^{1,2}$}
\affiliation{
$^1$\,JARA-FIT and 2nd Institute of Physics, RWTH Aachen University, 52074 Aachen, Germany, EU\\
$^2$\,Peter Gr\"unberg Institute (PGI-9), Forschungszentrum J\"ulich, 52425 J\"ulich, Germany, EU\\
$^3$\,National Institute for Materials Science,1-1 Namiki, Tsukuba, 305-0044, Japan\\
$^4$\,Department of Physics and Center for Advanced Materials and Nanotechnology, Lehigh University, Bethlehem, Pennsylvania 18015, USA\\
}
\begin{abstract}
Many-body effects resulting from strong electron-electron and electron-phonon interactions play a significant role in graphene physics.
We report on their manifestation in low $B$~field magneto-phonon resonances in high quality exfoliated single-layer and bilayer graphene encapsulated in hexagonal boron nitride.
These resonances allow us to extract characteristic effective Fermi velocities, as high as $1.20 \times 10^6$~m/s, for the observed ``dressed'' Landau level transitions, as well as the broadening of the resonances, which increases with Landau level index.
\end{abstract}
\maketitle

Scanning confocal Raman microscopy has emerged as a key tool for studying the unique properties of graphene.
In recent years, Raman spectroscopy has proven to be highly useful not only to identify graphene~\cite{ferrari2006,graf2007}, but also to extract information on local doping~\cite{ferrari2007,yan2007,pisana2007,stampfer2007,drogeler2014}, strain~\cite{mohr2010,huang2010,chacon2013}, and lattice temperature~\cite{calizo2007,balandin2008}.
Even more insights can be gained when combining Raman spectroscopy with magnetic fields.
In perpendicular $B$~fields, electronic states in graphene condense into Landau levels (LLs) which can interact with lattice vibrations.
Changing the magnetic field allows for both (i) tuning the Landau damping of the highest optical phonon mode at the $\Gamma$~point (G~mode) and (ii) shifting the Raman line when the G~mode is resonantly coupled to energetically matched LL transitions~\cite{ando2007,goerbig2007,kashuba2012,qiu2013}.
These coupled modes are known as magneto-phonon resonances (MPRs) and they provide a unique way to study electron-phonon interaction and many-body physics.
Although different MPRs have been measured in magneto-Raman experiments on graphene on graphite \cite{yan2010,faugeras2011,qiu2013} and multilayer graphene on SiC \cite{faugeras2009} and SiO$_2$ \cite{faugeras2012}, for exfoliated graphene only a single MPR at around $B=25$~T has been observed~\cite{kossacki2012,kim2013,leszczynski2014}.
While direct electronic LL-excitations for single- to penta-layer graphene have been recently studied for suspended graphene \cite{berciaud2014}, revealing interesting insights into Landau level physics, the interaction of Landau levels with the G~mode at low magnetic fields for exfoliated single-layer and bilayer graphene has not yet been observed and investigated.

\begin{figure}
\centering
\includegraphics[draft=false,keepaspectratio=true,clip, width=0.95\linewidth]{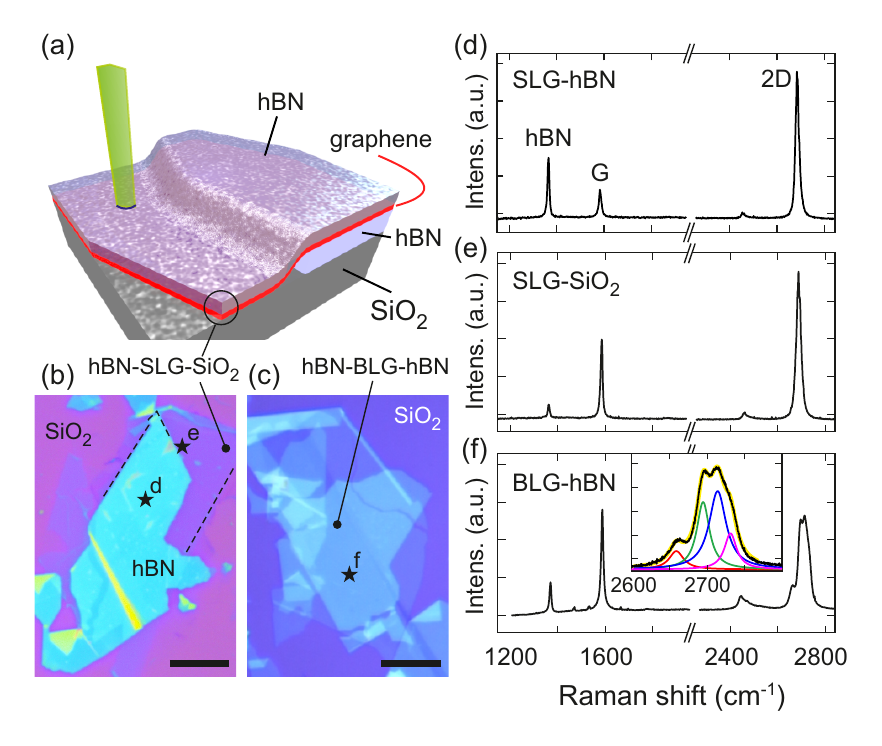}
\caption[FIG1]{
(a) Schematic illustration of a graphene-hBN heterostructure on SiO$_2$ and on hBN.
(b) and (c) Optical images of fabricated graphene, i.e. SLG (b) and BLG (c) samples.
The scale bars are 10~\textmu m.
The stars indicate where the Raman spectra in panels (d), (e) and (f) have been taken.
(d) and (e) Characteristic Raman spectra of graphene on hBN (d) and on SiO$_2$ (e).
(f) Characteristic Raman spectrum taken on the sample depicted in panel (c).
The inset shows a close-up of the 2D~peak.
The colored curves represent Lorentzian fits to the four sub-peaks with the yellow line representing the sum of the four fitted peaks. }
\label{fig1}
\end{figure}

Here we show magneto-Raman measurements on four different systems: (i) single-layer graphene (SLG) on SiO$_2$ covered with hexagonal boron nitride (hBN), (ii) SLG encapsulated in hBN, (iii) electrically contacted SLG encapsulated in hBN, and (iv) bilayer graphene (BLG) encapsulated in hBN.
The high quality of our hBN-graphene-hBN sandwich devices allows us to observe magneto-phonon resonances down to 2.1~T for SLG.
We are able to quantitatively study the influence of the electron-phonon interaction on the G~mode at low magnetic fields by pinning the charge carrier density in our samples.
In particular, we extract characteristic lifetimes for the observed Landau level excitations as well as high values of the effective Fermi velocities, a hallmark of electron-electron interaction effects, which appear to be LL transition and/or $B$~field dependent.
Finally, we report on MPRs in bilayer graphene.
Here the extracted effective Fermi velocities are lower, showing that many-body effects are less pronounced as compared to single-layer graphene.

A schematic illustration of a typical sample is shown in Figure~\ref{fig1}a, where a graphene flake is partly deposited on SiO$_{2}$ and partly on hBN.
The upper side of the graphene flake is completely covered with a second hBN flake.
This type of sample gives us invaluable capability to compare the material parameters of the \emph{same} graphene flake, where one surface is in direct contact with two kinds of substrates.
We employ a dry and resist-free fabrication method similar to Ref.~\nocite{wang2013}\citenum{wang2013}, where we pick-up an exfoliated graphene flake with an hBN flake and deposit it onto the hBN-SiO$_2$ transition area of the substrate.
This procedure has been shown to produce high-quality devices, as proven by transport measurements~\cite{wang2013,engels2014}.
An optical image of such a structure with single-layer graphene is shown in Figure~\ref{fig1}b.
In Figures~\ref{fig1}d and \ref{fig1}e we show two Raman spectra from the different substrate regions (see labeled regions and stars in Figure~\ref{fig1}b).
In both cases the characteristic hBN, G and 2D~peaks are observed.
The single Lorentzian shape of the 2D~line is characteristic for SLG.
The 2D~line also contains information on the underlying substrate.
For graphene encapsulated in hBN, we regularly find values of the full width at half maximum (FWHM) down to 16~cm$^{-1}$, while SiO$_{2}$ supported graphene shows values above 22~cm$^{-1}$.
An optical image of a bilayer graphene flake encapsulated in hBN is shown in Figure~\ref{fig1}c.
A corresponding Raman spectrum is shown in Figure~\ref{fig1}f revealing the typical 2D~line shape, which consists of four Lorentzians (see inset in Figure~\ref{fig1}f) \cite{ferrari2007,graf2007}.

\begin{figure*}
\centering
\includegraphics[draft=false,keepaspectratio=true,clip,width=1.0\linewidth]{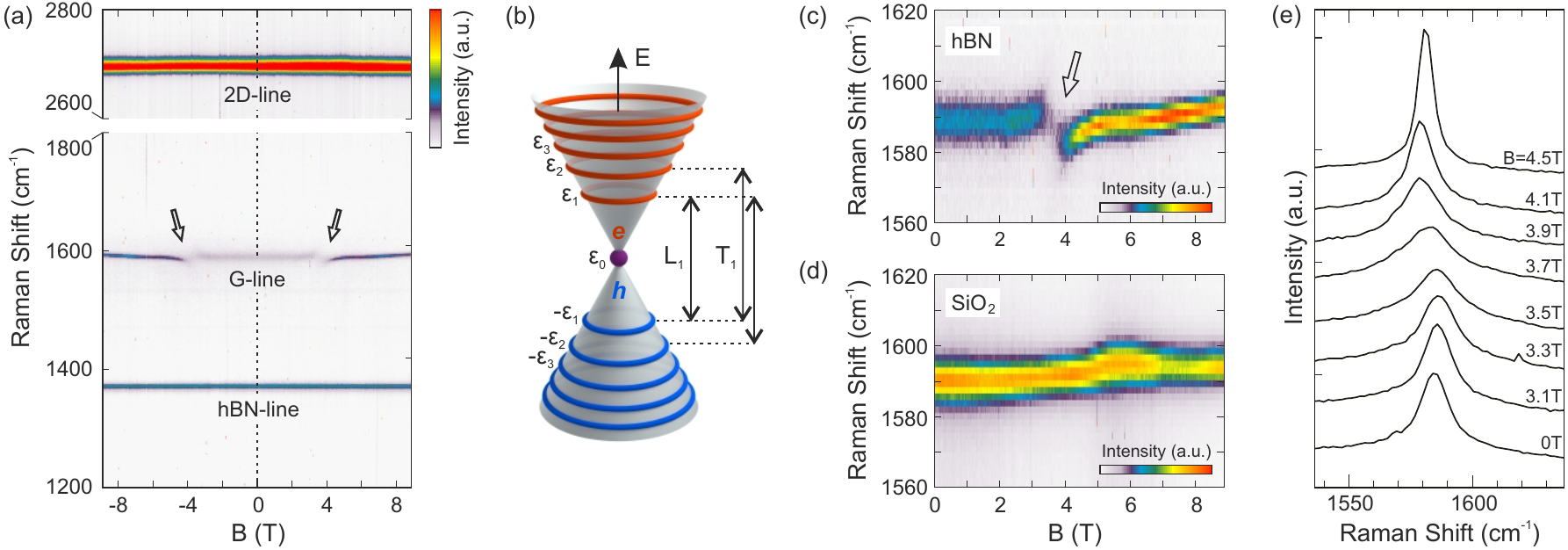}
\caption[FIG2]{ {Magneto-Raman spectroscopy on SLG on hBN and SiO$_2$.}
(a) Raman spectra plotted as a function of perpendicular magnetic field at $T=4.2$~K for a spot on the hBN-SLG-hBN sandwich area.
The G~line shows clear features around $B=\pm$3.7~T (see arrows).
(b) Schematic illustration of the linear graphene band structure, highlighting the different Landau levels and the inter-band Landau level transitions $T_1$ and $L_1$.
(c) Close-up of the G~line data shown in panel~(a).
(d) G~line as a function of magnetic field for a spot on hBN-graphene on SiO$_{2}$.
(e) Individual Raman spectra for different values of the magnetic field on the hBN-graphene-hBN area.
The Raman spectra are vertically offset for clarity.
The G~line can be well described by a single Lorentzian in all cases. }
\label{fig2}
\end{figure*}

For the low temperature magneto-Raman measurements we employ a commercially available confocal Raman setup, allowing us to perform spatially-resolved experiments at a temperature of 4.2~K and magnetic fields of up to 9~T.
We use an excitation laser wavelength of 532~nm with a spot diameter on the sample of around 500~nm.
For detection, we use a CCD spectrometer with a grating of 1200~lines/mm.
All measurements in this work are performed with linear laser polarization.

In Figure~\ref{fig2}a we show a color-encoded two-dimensional plot of the Raman intensity as a function of magnetic field and Raman shift for a spot on the area of SLG encapsulated in hBN.
For the hBN and 2D~line we observe only a weak $B$~field dependence (see Supplementary Information), in agreement with earlier studies~\cite{faugeras2010}.
The G~line, however, shows significant $B$~field dependence, which result from the resonant coupling of the G~mode to inter-band LL transitions whose energies are given by $T_{n} = \varepsilon_{n+1} + \varepsilon_{n}$, where $\varepsilon_{n} = v_{\mathrm{F}} \sqrt{2 \hbar e B n}$ is the absolute value of the energy of the $n$th Landau level.
In particular, the feature at $B \approx \pm 3.7$~T (see arrows in Figure~\ref{fig2}a) can be attributed to the resonant coupling of the $T_1$-transition (see Figure~\ref{fig2}b) and the G~mode.
A close-up around the G~line is shown in Figure~\ref{fig2}c, with the arrow highlighting the $T_1$-MPR.
In contrast to the magneto-Raman spectra taken on SLG encapsulated in hBN, the spectra obtained on SLG deposited on SiO$_2$ do not show any resonant behavior as highlighted in Figure~\ref{fig2}d, indicating the strong influence of the substrate material on the properties of graphene.
We therefore focus our discussion on the hBN-SLG-hBN data.
Interestingly, a close inspection of the individual Raman peaks at different $B$ fields (see Figure~\ref{fig2}e) reveals that the G~line can be well described by a single Lorentzian at all $B$~fields even across the resonance.

\begin{figure}
\centering
\includegraphics[draft=false,keepaspectratio=true,clip,width=0.90\linewidth]{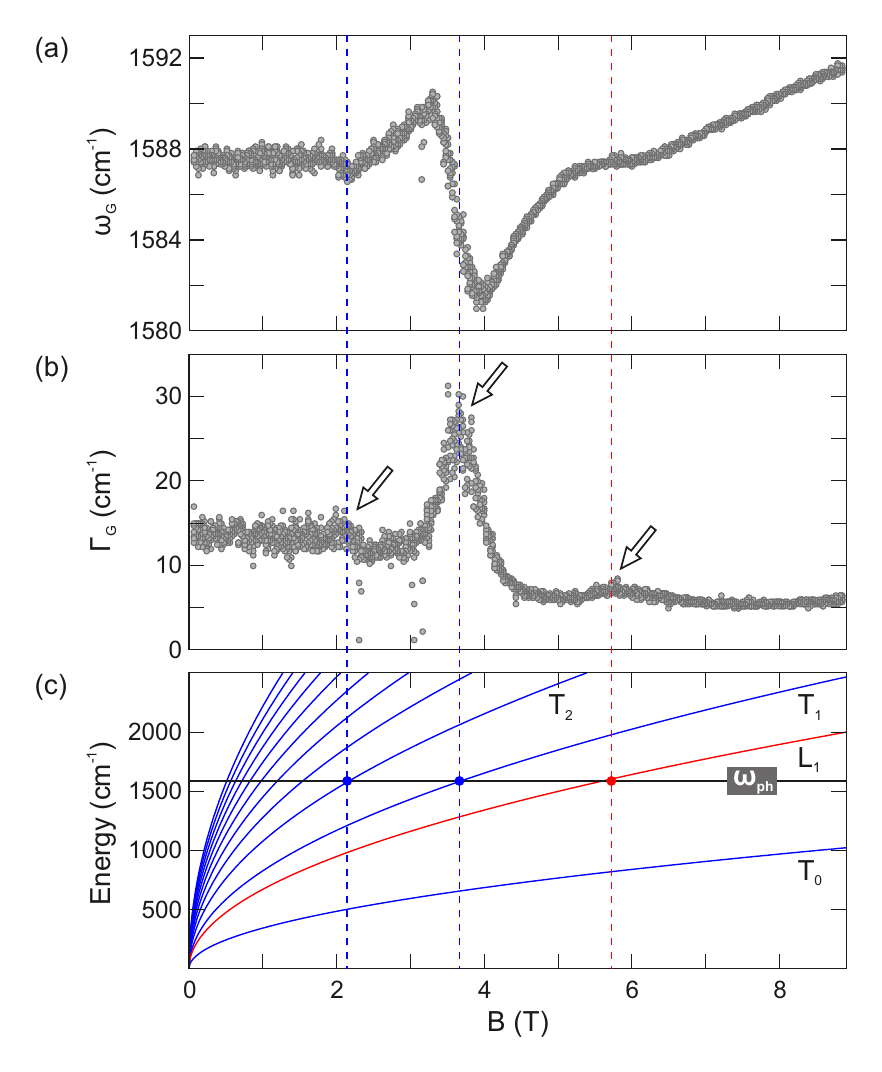}
\caption[FIG3]{ {Magneto-phonon resonances on hBN-SLG-hBN heterostructure.}
(a) and (b) G~line position (a) and G~line FWHM (b) as functions of $B$ field as obtained from Lorentzian fits to the data.
The arrows highlight the three visible MPRs.
(c) Inter band Landau level transition energies $T_n$ (blue lines) and $L_1$ (red line) as a function of magnetic field.
The black line represents the G~mode frequency at zero magnetic field, $\omega_{\mathrm{ph}}$.
For the plot, effective Fermi velocities of $v_{\mathrm{F}} = 1.17 \times 10^{6}$~m/s (blue curves) and $v_{\mathrm{F}} = 1.12 \times 10^{6}$~m/s (red curves) and $\omega_{\mathrm{ph}} = 1587$~cm$^{-1}$ are used. }
\label{fig3}
\end{figure}

\begin{figure}
\centering
\includegraphics[draft=false,keepaspectratio=true,clip,width=0.95\linewidth]{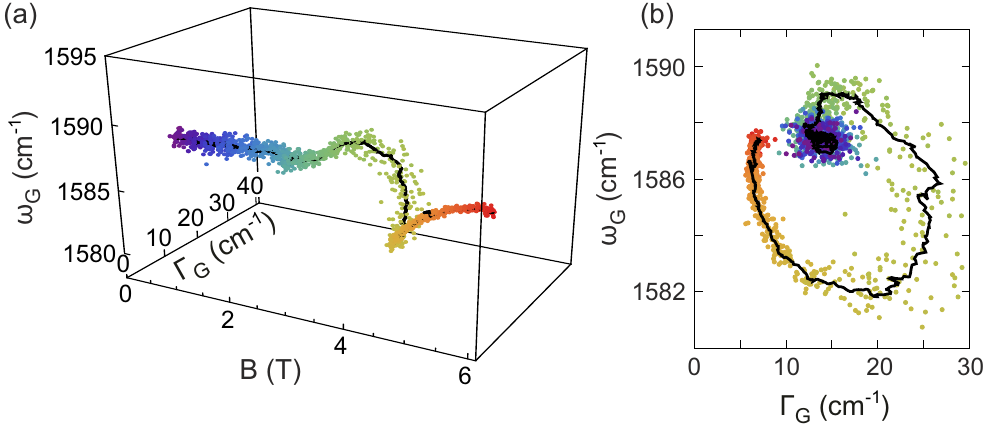}
\caption[FIG4]{
(a) Three-dimensional representation of the data shown in Figure~\ref{fig3}.
(b) $\omega_G$ versus $\Gamma_G$.
The $B$~field value of each point is color encoded and corresponds to the points shown in panel (a). }
\label{fig4}
\end{figure}

In Figures~\ref{fig3}a and \ref{fig3}b, we show the evolution of the measured position $\omega_G$ and width $\Gamma_G$ of the G~line with $B$~field.
We observe three resonances at around 2.1~T, 3.7~T, and 5.8~T (see e.g. arrows in Figure~\ref{fig3}b).
At these values, the energies of the LL transitions $T_2$, $T_1$, and $L_1$ (see below) match
the energy of the G mode phonon at zero $B$ field $\hbar \omega_{\mathrm{ph}}$.
This is highlighted in Figure~\ref{fig3}c, where we show the dependence of $T_n$ and $L_1$ on $B$~field (see also vertical dashed lines in Figure~\ref{fig3}).
The resonant behavior of the most prominent feature at 3.7~T can be further visualized by plotting the data in a three-dimensional representation (Figure~\ref{fig4}a) and as an Argand diagram (Figure~\ref{fig4}b).
When projected on the $\omega_{\mathrm{G}}$-$\Gamma_{\mathrm{G}}$ plane (Figure~\ref{fig4}b) the resonance appears as a circle.
This behavior is a hallmark of a special case of the dynamics of a non-Hermitian two level system \cite{hernandez2006,rotter2009,graefe2010}.
The horizontal width of the Argand circle $\Delta\Gamma_G\simeq 20-25~$cm$^{-1}$ depends on the strength of the electron-phonon interaction at the $T_1$ level crossing as well as on the charge carrier density $n_{\mathrm{el}}$, which might block LL excitations due to the Pauli principle (see Supplementary Information), and the lifetimes of the involved states.

\begin{figure*}
\centering
\includegraphics[draft=false,keepaspectratio=true,clip,width=1.0\linewidth]{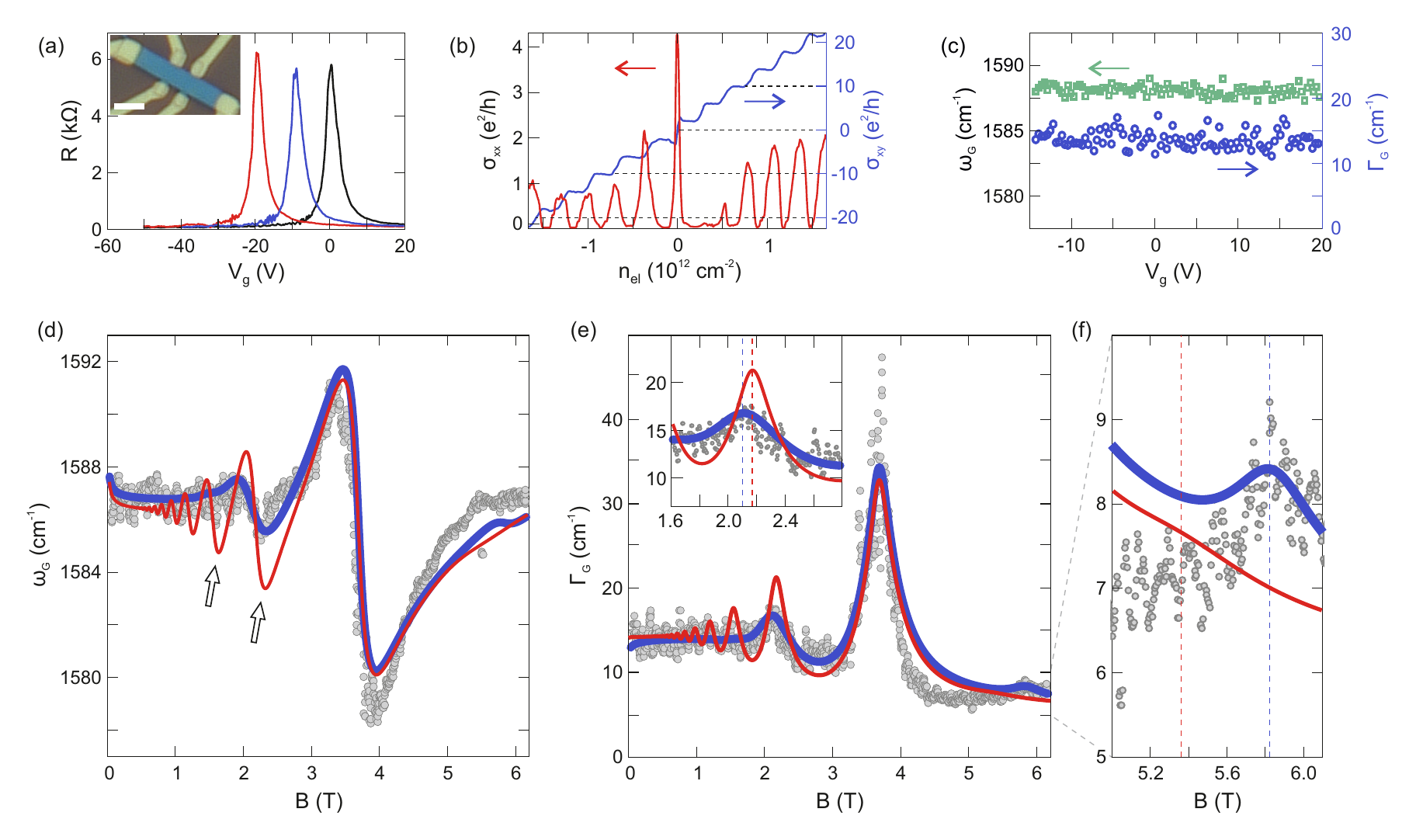}
\caption[FIG5]{{ Transport and magneto-Raman measurements of a contacted hBN-graphene-hBN heterostructure.}
(a) Longitudinal sample resistance as a function of back gate voltage after photo doping at $V_{\mathrm{g}} = -20$~V (red), {-10}~V (blue) and 0~V (black).
The inset shows an optical image of the device, with the scale bar representing 5~\textmu m.
(b) Longitudinal conductivity $\sigma_{xx}$ (red) and Hall conductivity $\sigma_{xy}$ (blue) plotted as a function of charge carrier density at $B$ = 3.7 T. The typical half-integer quantum Hall plateaus for graphene are observed.
(c) G~line position (blue) and G~line FWHM (green) as a function of $V_{\mathrm{g}}$.
(d) and (e) $\omega_G$ and $\Gamma_G$ as functions of magnetic field recorded with $V_{\mathrm{g}}$ set to 0~V.
The data points are extracted from Lorentzian fits similar to Figure~3.
A running average of 25~mT was used.
The red lines represent a theoretical model which uses constant $v_{\mathrm{F}} = 1.17 \times 10^6$~m/s and constant $\gamma_{\mathrm{el}} = 160$~cm$^{-1}$ for all resonances.
The blue lines represent the theoretical model, where different effective Fermi velocities ($v_{\mathrm{F},T_1} = 1.17$~m/s, $v_{\mathrm{F},T_2} = 1.19$~m/s, $v_{\mathrm{F},L_1} = 1.12$~m/s) and electronic broadening parameters ($\gamma_{\mathrm{el},T_1} = 160$~cm$^{-1}$, $\gamma_{\mathrm{el},T_2} = 270$~cm$^{-1}$, $\gamma_{\mathrm{el},L_1} = 80$~cm$^{-1}$) are employed for the $T_1$-, $T_2$- and $L_1$-resonances.
The $T_n$ for $n \geq 3$ are modeled with $\gamma_{\mathrm{el},T_n} = 400$~cm$^{-1}$ to fully suppress them, while $\gamma_{\mathrm{el},T_0}$ was set to 20~cm$^{-1}$.
The other parameters were fixed to $\omega_{\mathrm{ph}} = 1586$~cm$^{-1}$,  $\gamma_{\mathrm{ph}} = 5.5$~cm$^{-1}$, $\lambda = 4 \times 10^{-3}$, and $\lambda_L = 0.015 \lambda$.
The inset of panel~(e) and panel~(f) show close-ups of the $T_2$- and $L_1$-MPRs, respectively, highlighting the differences between the blue and red traces. }
\label{fig5}
\end{figure*}

In order to separate the influence of these effects and to gain quantitative values, an independent control of the charge carrier density is desirable.
This is achieved by repeating the MPR-experiment on an electrically contacted hBN-graphene sandwich device.
Although different processing steps, including electron beam lithography, reactive ion etching, and metal (Cr/Au) evaporation, are needed~\cite{wang2013,engels2014b}, the sample quality allows the observation of MPRs of similar quality and at similar $B$ fields as for the unprocessed sample in Figure~\ref{fig3}.
The inset of Figure~\ref{fig5}a shows the investigated device.
A four-terminal back gate characteristic of the graphene resistance is shown as a black trace in Figure~\ref{fig5}a and a carrier mobility of around $\mu \approx 4 \times 10^4$~cm$^2$/(Vs) and a width of the conductance minimum~\cite{couto2014} of $n_{\mathrm{el}}^* \approx 7 \times 10^{10}$~cm$^{-2}$ are extracted from these data (see Supplementary Information).
The high quality of our sample is also seen in the quantum Hall measurement (see Figure~\ref{fig5}b), where already at $B = 3.7$~T well-separated LLs are established, which is in good agreement with our observation of the $T_1$-MPR.
To pin the charge carrier density close to the CNP, we make use of the recently reported photo-doping effect in graphene-hBN-heterostructures \cite{ju2014}.
When shining light of sufficient intensity on a gated graphene-hBN sandwich device, nitrogen vacancies or carbon impurities in the hBN can get charged up to the point where they completely screen the electric field due to the applied back-gate voltage.
As a consequence, laser illumination of a graphene-hBN sandwich device pins the carrier density in graphene very close to the CNP, independent of the applied gate voltage.
This happens on time scales much faster than the Raman acquisition time, thanks to a laser power of 2~mW at a spot size of 0.25~\textmu m$^2$, which corresponds to an intensity that is a factor $\sim$10$^3$ higher than that used in Ref.~\nocite{ju2014}\citenum{ju2014}.
The pinning of the carrier density results in the fact that the position and the line width of the G~line become essentially independent of the back-gate voltage, as shown in Figure~\ref{fig5}c.
When turning off the laser, the CNP remains pinned to the last value of the gate voltage that was applied before.
This memory effect, also called photo-induced doping~\cite{ju2014}, is shown in Figure~\ref{fig5}a, where the red and blue traces represent two back gate characteristics that are shifted by the photo-induced doping effect, but are otherwise nearly unmodified.
Importantly, it has been shown that the width $n_{\mathrm{el}}^*$ of the conductance minimum is independent of the photo-doping value~\cite{ju2014}.
This quantity is thus a good estimate for the upper limit of the pinned charge-carrier density in graphene under Raman measurements.
For the investigated sample we find $|n_{\mathrm{el}}| <  n_{\mathrm{el}}^* \approx 7 \times 10^{10}$~cm$^{-2}$ (see Supplementary Information).

Having knowledge of the charge carrier density, we can now follow Ando \cite{ando2007} and Goerbig {\it et al.} \cite{goerbig2007} to quantitatively investigate the low $B$~field MPRs (see gray data points in Figures~\ref{fig5}d and \ref{fig5}e).
It has been shown that the dependence of the position $\omega_G$ and the width $\Gamma_G$ of the G~line as a function of the magnetic field can be well understood in terms of the renormalization of the phonon propagator due to electron-phonon interaction.
The phonon self-energy $\Pi(\omega)$ for (nearly) zero doping is given by

\begin{equation*}
\Pi(\omega) = e v_{\mathrm{F}}^2 B \left( \lambda \sum_{n=0}^N \xi(T_n;\omega) + \lambda_L \sum_{n=1}^N \xi(L_n;\omega) \right),
\label{eq1}
\end{equation*}
where
\begin{equation*}
\xi(\varepsilon;\omega) = \frac{2(\varepsilon - i\hbar\gamma_{\mathrm{el},\varepsilon}/2)}{(\hbar \omega)^2 - (\varepsilon - i\hbar\gamma_{\mathrm{el},\varepsilon}/2)^2} + \frac{2}{\varepsilon - i\hbar\gamma_{\mathrm{el},\varepsilon}/2}.
\label{eq2}
\end{equation*}

In the equations above, $L_n = 2\varepsilon_n$ represents the $\Delta n = 0$ inter-band LL transition energies, while $T_n$ denotes the energies of the $\Delta n = 1$ inter-band LL transitions as defined above (see also Figure~\ref{fig2}b).
The according $\gamma_{\mathrm{el},T_n}$ and $\gamma_{\mathrm{el},L_n}$ account for the finite lifetimes of the different LL excitations and
$\lambda$ and $\lambda_L$ denote the coupling constants of the $T_n$- and $L_n$-transitions to the G~mode, respectively.
The coupling of the $L_n$-transitions to the G mode, which has also been observed in previous studies\cite{faugeras2011,kuhne2012}, is only due to higher order processes \cite{ando2007}, generally making $\lambda_L \ll \lambda$.
The $L_1$~transition, in particular, has to be included as it is needed to account for the resonance at around 5.8~T.
Since the charge carrier density in our sample is low enough not to affect the strength of the resonances, we do not need to take into account the filling factor dependence of $\Pi(\omega)$.
Furthermore, near zero doping, all intra-band LL excitations are blocked by the Pauli principle.

The above expression for the phonon self-energy can finally be used to find the pole of the renormalized phonon propagator by solving the equation $\omega^2-\omega_{\mathrm{ph}}^2-2\omega_{\mathrm{ph}}\Pi(\omega)=0$.
Its roots are related to the position $\omega_G$ and FWHM $\Gamma_G$ of the G~mode via $\omega = \omega_G - i\Gamma_G/2$.
To account for non-electronic broadening effects, we make the replacement $\omega_{\mathrm{ph}} \to \omega_{\mathrm{ph}} -i\gamma_{\mathrm{ph}}/2$, with $\gamma_{\mathrm{ph}}$ being the broadening of the phonon due to non-electronic processes
and $\omega_{\mathrm{ph}}$ the G~mode frequency at zero $B$ field.

In Figures~\ref{fig5}d, \ref{fig5}e and \ref{fig5}f we show the comparison between calculation and experiment. The strength of the MPRs is comparable to that on the unprocessed samples, showing that our etching and contacting techniques do not reduce the device quality.
The red traces show the theoretical result taking a constant Fermi velocity of $v_{\mathrm{F}} = 1.17 \times 10^6$~m/s and a constant broadening parameter $\gamma_{\mathrm{el}} = \gamma_{\mathrm{el},T_n} = \gamma_{\mathrm{el},L_n} = 160$~cm$^{-1}$.
These two parameters were chosen such that the $T_1$-MPR is described well.
Values for the other parameters are obtained as follows.
While the G mode phonon frequency at $B = 0$~T, $\omega_{\mathrm{ph}} = 1586$~cm$^{-1}$, can be directly extracted from the data, the non-electronic phonon broadening $\gamma_{\mathrm{ph}} = 5.5$~cm$^{-1}$ can be extracted from the residual $\Gamma_G$ at higher magnetic fields (e.g. at 8~T) where no LL transitions are energetically matched with the phonon mode.
The electron-phonon coupling parameter $\lambda = 4 \times 10^{-3}$ is fixed by the value of $\Gamma_G$ at very low magnetic fields as it has to guarantee the right amount of Landau damping.
The extracted $\lambda$ is in good agreement with values reported by other groups \cite{yan2007,faugeras2009,goler2012,kuhne2012,kim2013,qiu2013}.
For the phenomenologically introduced $\lambda_L$ we used a value of $\lambda_L = 0.015 \, \lambda$.
A constant $\gamma_{\mathrm{el}}$ for all LL excitations evidently overestimates the strength of the MPRs at low magnetic field values (see arrows in Figure 5d).
Moreover, the positions of the resonances do not match between theory and experiment if a constant Fermi velocity is used for all LL excitations (see vertical dashed lines in Figures~\ref{fig5}e and \ref{fig5}f).
In particular, the $L_1$-MPR appears in the calculation at a lower magnetic field value than in the experimental data.
We thus phenomenologically modify the model and assign different widths and effective Fermi velocities (following Shizuya~\cite{shizuya2010}) to the three visible MPRs (see caption of Figure~\ref{fig5} and Table~\ref{tab1}) and end up with the blue traces in Figures~\ref{fig5}d-5f, which are in better quantitative agreement with the experimental data.

Interestingly, we observe that the extracted effective Fermi velocities are significantly higher compared to earlier MPR and infrared measurements on graphene and graphene related systems \cite{jiang2007,yan2007,faugeras2009,goler2012,kuhne2012,kim2013,qiu2013}.
We attribute the high values of $v_{\mathrm{F}}$ to many-body effects which arise since our samples consist of exfoliated graphene single-layers with low doping and low doping fluctuations.
So far only suspended SLG showed similarly high values \cite{berciaud2014,footnote1}.
The $v_{\mathrm{F}}$~values we extracted from the measurement of MPRs on 13 spots on three different samples are displayed in Table~\ref{tab1}.
The errors on $v_{\mathrm{F}}$ are below $0.01 \times 10^{6}$~m/s.
The extracted values indicate that $v_{\mathrm{F}}$ decreases with $B$~field and might also depend on the transition as has been predicted by theory \cite{iyengar2007,shizuya2010,shizuya2011}.

Finally, we find that the widths of the LL excitations are strongly dependent on the LL indices of the involved states
(see last column in Table 1).
The electronic lifetime decreases with increasing LL index, resulting in strong suppression of all $T_n$ for $n \geq 3$.
This observation might be attributed to the additional decay channels arising from an increased number of energetically lower and unoccupied LLs for excitations with higher $n$.
However, we cannot gain independent insights on the $B$ field dependence of the LL broadening~\cite{jiang2007} as we are limited to extract $\gamma_{el}$ for each transition from the corresponding MPR.

\begin{table*}
\begin{centering}
\begin{tabular}{ccccc}
\hline
Sample: & \quad $\;\!$ \#A \quad $\;\!$ & \quad $\;\!$ \#B \quad $\;\!$ & \quad $\;\!$ \#C \quad $\;\!$ & \quad $\;\!$ \#C \quad $\;\!$ \tabularnewline
& \multicolumn{3}{c}{$v_{\mathrm{F}}$~($\times 10^6$~m/s)} & $\gamma_{\mathrm{el}}$~(cm$^{-1}$) \tabularnewline
\hline
$T_2$~($\sim$~2.1~T) & 1.20 & 1.19 & 1.20 & 270 \tabularnewline
$T_1$~($\sim$~3.7~T) & 1.17 & 1.16 & 1.17 & 160 \tabularnewline
$L_1$~($\sim$~5.8~T) & 1.13 & 1.13 & 1.12 & 80 \tabularnewline
\hline
\end{tabular}
\par\end{centering}
\centering{}
\caption{ Effective Fermi velocities extracted from the position of the MPRs on three different samples.
\#A: sample shown in Figure~\ref{fig1}a.
\#B: second, non-contacted hBN-graphene-hBN sample.
\#C: contacted hBN-graphene sandwich characterized in Figure~\ref{fig5}.
The values in the first three columns contain the mean values of the effective Fermi velocities as measured on a varying number of spots per sample.
The last column contains the LL excitation widths used in the calculation. }
\label{tab1}
\end{table*}

\begin{figure}
\centering
\includegraphics[draft=false,keepaspectratio=true,clip,width=0.90\linewidth]{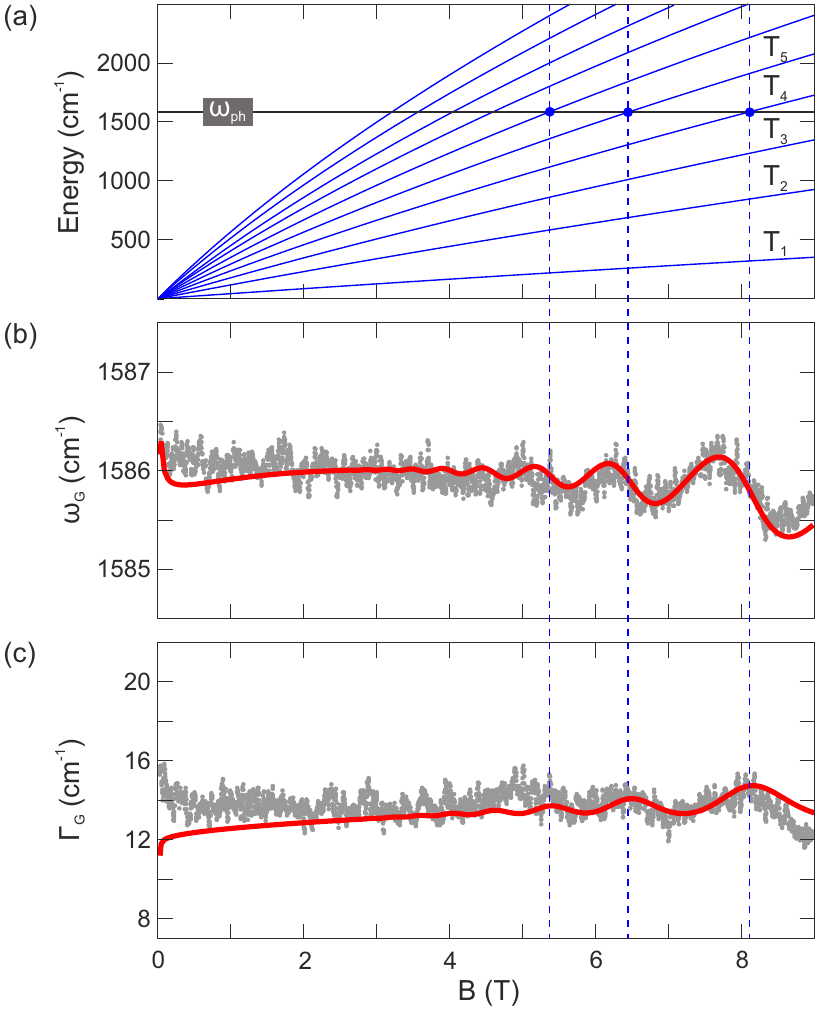}
\caption[FIG6]{ {Magneto-phonon resonances in bilayer graphene encapsulated in hBN.}
(a) inter-band Landau level transition energies $T_n$ (blue lines) as a function of $B$~field.
The black line represents the phonon frequency at zero $B$~field.
(b) and (c) $\omega_G$ and $\Gamma_G$ as functions of $B$~field.
The data points are extracted from Lorentzian fits to data taken on the sample shown in Figure~\ref{fig1}c.
A running average of 25~mT was used.
The red lines are computed using equation Eq.~S2 (see Supplementary Information) with the parameters given in the text. }
\label{fig6}
\end{figure}

Our fabrication technique, yielding samples with high carrier mobility and low doping, can also be applied to bilayer graphene.
This allows us to investigate MPRs in exfoliated BLG, which could not have been studied earlier.
The most relevant LL transitions in BLG are inter-band excitations in the lower subbands, which we also denote by $T_n = \varepsilon_{n+1} + \varepsilon_n$, where $\varepsilon_n$ now denotes the energy of a Landau level in the lower subband of BLG (see Supplementary Information) and $n \geq 1$.
Their evolution with $B$~field is shown in Figure~\ref{fig6}a.
As for the case of SLG, we fit single Lorentzians to every Raman spectrum.
The extracted values of $\omega_G$ and $\Gamma_G$ as a function of $B$~field are shown in Figures~\ref{fig6}b and \ref{fig6}c, respectively.
Two resonances at around 6.4~T and 8.1~T can be well identified, which can be attributed to the LL transitions $T_5$ and $T_4$, respectively.
The red curves are calculated using Eq.~S2 (see Supplementary Information), with $v_{\mathrm{F}} = 1.07 \times 10^6$~m/s, $\lambda=3.5 \times 10^{-3}$, $\gamma_{\mathrm{ph}} = 6.5$~cm$^{-1}$, $\omega_{\mathrm{ph}}= 1585.9$~cm$^{-1}$ and $\gamma_{\mathrm{el}} = 270$~cm$^{-1}$.
Overall, the magnitudes of the resonances below 9~T are significantly lower as compared to the case of single-layer graphene (see Figure~\ref{fig3}), which, however, is in good agreement with our calculation.
Interestingly, in contrast to single-layer graphene, a constant Fermi velocity is sufficient to describe our data.
Moreover, compared to SLG, the Fermi velocity is reduced, indicating that electron-electron interaction effects play a minor role in bilayer graphene.
However, as we do not know the charge carrier doping in our bilayer sample, a suppression of many-body effects might also be due to finite charge carrier density.
It is possible to describe both resonances with the same LL exciton broadening.

In summary, we investigated the $B$~field dependence of the G~line phonon renormalization of exfoliated single-layer and bilayer graphene for low magnetic fields.
For single-layer graphene we have compared the substrate influence of SiO$_2$ and hBN.
While no distinct features could be observed on SiO$_2$, the Raman spectra on hBN showed distinct magneto-phonon resonances.
For a quantitative discussion of the parameters involved in the coupling of Landau level transitions to the Raman G~mode we investigated an electrically contacted graphene sample to pin the charge carrier density.
We found high effective Fermi velocities of up to $1.20 \times 10^6$~m/s, which varied for the three observed LL excitations.
Due to the low charge carrier density, we attribute this finding to many-body contributions to the energy of the LL excitons.
We show that Landau level excitations with higher Landau level index have reduced lifetimes, possibly linked to the existence of an increased number of intermediate LL states that lead to additional decay channels.
Finally, we observed MPRs in exfoliated bilayer graphene.
Here, a constant Fermi velocity and a single Landau level excitation width for all resonances is sufficient to reach good agreement with theory.
Our work paves the way towards a more refined understanding of electronic many-body effects in the presence of magnetic fields in single-layer and bilayer graphene and can be easily extended to other two-dimensional materials.

\section{acknowledgement}

We thank T. Khodkov, C. Volk, P. Kaienburg and C. B{\"o}defeld (from attocube systems AG) for fruitful discussions.
Support by the Helmholtz Nanoelectronic Facility (HNF), the DFG, the ERC (GA-Nr. 280140) and the EU projects Graphene Flagship (contract no. NECT-ICT-604391), are gratefully acknowledged.



\providecommand*\mcitethebibliography{\thebibliography}
\csname @ifundefined\endcsname{endmcitethebibliography}
  {\let\endmcitethebibliography\endthebibliography}{}

\end{document}